\documentclass[12pt]{iopart}

\usepackage{hyperref}
\usepackage{graphicx}
\usepackage{rotating}
\usepackage{array}
\usepackage{amssymb}
\usepackage{tikz}

\def\v#1{{\bf#1}}
\def\be{\begin{equation}}
\def\ee{\end{equation}}
\def\bea{\begin{eqnarray}}
\def\eea{\end{eqnarray}}

\def\ie{{\it i.e.\,}}
\def\etal{{\it et al.   }}

\def\<{\langle}
\def\>{\rangle}

\begin{document}

\title{Canonical Transformations in Crystals}
\author{E. Sadurn\'i }

\address{Instituto de F\'isica, Benem\'erita Universidad Aut\'onoma de Puebla,
Apartado Postal J-48, 72570 Puebla, M\'exico}

\eads{ \mailto{sadurni@ifuap.buap.mx}}

\begin{abstract}

\noindent
The space representations of linear canonical transformations were studied by Moshinsky and Quesne in 1972. For a few decades, the bilinear hamiltonian remained as the only exactly solvable representative for such problems. In this work we show that the Mello-Moshinsky equations can be solved exactly for a class of problems with discrete symmetry, leading to exact propagators for Wannier-Stark ladders in one and two dimensional crystals. We give a detailed study for a particle in a triangular lattice under the influence of a time-dependent electric field. A more general set of Mello-Moshinsky equations for arbitrary lattices is presented.

\end{abstract}


\maketitle




\section{Introduction}

It is well known that the study of quantum-mechanical representations of canonical transformations was initiated with Bargmann's work \cite{bargmann}, although Dirac's book \cite{dirac} already gave a hint to this important subject. In 1972 the problem was revisited by Moshinsky and Quesne \cite{moshandquesne}, who found that the space representations of linear canonical transformations in arbitrary dimensions were given by gaussian kernels. It was then clear that the propagator associated to any bilinear hamiltonian -- in space and momentum -- could be obtained explicitly with group theoretical methods, in this case the symplectic group. Notably, the harmonic oscillator with time-dependent mass and frequency belongs to this category \cite{3.2}. 

For four decades, the bilinear hamiltonian remained as the only example for which the transformation equations -- also known as the Mello-Moshisnky (MM) equations \cite{mosh} -- could be solved analytically in terms of elementary functions \footnote{An amusing example involving a quadratic canonical transformation and Airy functions comes from the time-energy variables $(t,H)$ of a particle in a linear potential \cite{moshLASP}. Such a transformation does not connect $(x,p)$ with $(t,H)$ continuously, and shall not be considered in our discussions.}. 

In this contribution we show that new solutions for the representations of evolution maps can be obtained in the context of tight-binding models. In particular, we shall find the propagator of a particle on a lattice under the influence of a time-dependent homogeneous field. The corresponding MM equations shall be written as recurrence relations on the {\it discrete\ }coordinates of the system, finding thereby that the kernels can be written in terms of Bessel functions or their multi-index generalizations \cite{1.3}.

From the physical point of view, we may also point out that the study of the evolution of wave packets in crystals has a long history of its own, which started with Bloch's work in 1928 \cite{bloch}. Solid state physics and, quite recently, matter waves in lattice traps \cite{ibloch} serve as a motivation to our work.  

In connection with canonical variables in crystals, we underscore the fact that such operators {\it can\ }be defined by a careful construction in terms of position operators and crystal translations \footnote{Interesting possibilities were provided in \cite{smirnov}. In this paper we follow a different approach.}. In section 2 we take care of this point by providing the canonical and non-canonical maps that set the foundations in further developments. In section 3 we formulate the discrete MM equations for canonical transformations in one-dimensional lattices and obtain explicitly the propagator of a model hamiltonian, namely the particle in discrete space with time-dependent hopping parameters and a time-dependent homogeneous field. Section 4 deals with two-dimensional generalizations of the problem: triangular and square lattices. There, we obtain the corresponding MM equations and their solution. In section 5 we formulate the problem using tight-binding lattices in arbitrary dimensions and coordination numbers. We conclude briefly in section 6.

\section{Independent vs. canonical variables in one-dimensional lattices}

The study of canonical transformations as a function of time is usually based on the invariance of $\left[x,p \right]$. We show first that a canonical pair in discrete variables can be constructed. Then we propose another pair which is non-canonical, but whose commutator transforms canonically; through such a pair we will be able to perform computations easily in further sections. A schematic view of the procedure is given in figures \ref{diag1} and \ref{diag2}.  

\subsection{A canonical pair in discrete space}
The first question arising in the context of crystalline structures is whether one can define a momentum operator canonically conjugate to a discrete position. The concept of derivative is obviously not applicable here, but a path offers itself through the central discretization of the derivative. Let us define our units in the form $\hbar=1$ and denote by $a$ the lattice spacing parameter with units of length. Then we have the following example of canonical operators $\xi, \pi$ in terms of the usual continuous-space operators $x$ and $p= - i\partial / \partial x$:

\bea
\xi \equiv \frac{1}{2} \left\{\sec(ap) , x \right\}, \qquad \pi \equiv \frac{1}{a} \sin(ap).
\label{1}
\eea 
One can easily check that $a \rightarrow 0$ leads to an appropriate continuous limit and that the new operators $\xi, \pi$ satisfy the canonical commutator
\bea
\left[\xi,\pi \right] = i \qquad \Longleftrightarrow \qquad \left[x,p \right] = i.
\label{2}
\eea
Although the map (\ref{1}) is not invertible --we shall discuss this later on-- its functional form guarantees that unitary transformations $UxU^{\dagger}, UpU^{\dagger}$ representing arbitrary canonical maps \footnote{We shall focus on unitary representations of canonical transformations, embracing probability conservation as our guiding principle.} induce canonical transformations on the pair $\xi, \pi$. In fact, the transformation rule (\ref{1}) can be written as a power series of $x$ and $p$:
\bea
\xi = \sum_{n=0}^{\infty}\frac{(ia)^{2n}E_{2n}}{2(2n)!} \left\{ p^{2n} , x \right\}, \qquad \pi = \sum_{n=0}^{\infty} \frac{a^{2n}}{(2n+1)!} p^{2n+1},
\label{3}
\eea
where $E_{2n}$ denotes the Euler number. Therefore one can easily show that
\bea
\xi' \equiv \xi \left(  U x U^{\dagger} ,  U p U^{\dagger} \right) = U \xi U^{\dagger},
\label{3.9}
\eea
\bea
\pi' \equiv \pi \left(  U x U^{\dagger} ,  U p U^{\dagger} \right) = U \pi U^{\dagger}. 
\label{4}
\eea
Moreover, the meaning of these operators can be found by their application to localized (or atomic) states. Let us denote such sates by $|n\>$, where $n \in \v Z$ is the site number and $\psi_n(x)=\<x|n\>$ represents the Wannier function \footnote{Only one band is considered for simplicity. The band index of the Wannier function shall be omitted.} around site $n$. We have

\bea
\xi \psi_n (x) &=& \sum_{m=0}^{\infty} (-)^m \left(2x-\left[2m+1 \right]a \right) \psi_n \left(x-\left[2m+1 \right]a \right) \nonumber \\ &=& \sum_{m=0}^{\infty} (-)^m \left(2x-\left[2m+1 \right]a \right) \psi_{n+2m+1} \left(x \right),
\label{4.1}
\eea
\bea
 \pi \psi_n(x) &=& \frac{\psi_n(x+ a) - \psi_n(x- a)}{2 i a} \nonumber \\ &=&  \frac{\psi_{n-1}(x) - \psi_{n+1}(x)}{2 i a}.
\label{4.2}
\eea
By virtue of these relations and the transformations $(\ref{3.9}), (\ref{4})$, the study of canonical transformations in crystals through discrete variables is justified; the action of $\xi$ and $\pi$ on functions of a continuous variable $x$ can be written explicitly as an operation on the discrete variable $n$, as indicated by the following vector representation of (\ref{4.1}) and (\ref{4.2}):

\bea
\xi |n\>&=& a \sum_{m=0}^{\infty} (-)^m \left[ 2(n-m) - 1 \right] |n+2m+1\> ,
\label{4.3}
\eea
\bea
 \pi | n \>  &=&  \frac{1}{2 i a} | n-1 \> -  \frac{1}{2 i a} | n+1 \>.
\label{4.4}
\eea
 Finally, through a map such as (\ref{1}), it is fair to restrict our considerations to the invariance of the commutator $\left[\xi,\pi \right]$ rather than $\left[x,p \right]$. Thus, our first question can be answered.

\subsection{A convenient pair of non-canonical variables}

For practical reasons one might be interested in non-canonical variables whose transformations are more amenable; according to (\ref{4.1}) the operator $\xi$ acts non-locally on lattice functions, giving complicated expressions for the matrix elements $\<n| \xi |m\>$. Other operators given by arbitrary functions $N(x,p)$ and $T(x,p)$ might allow a simpler analysis of canonical transformations by means of a covariant $\left[N, T \right]$. A careful explanation of this process is given in \cite{moshLASP}, where it was pointed out that the elements of $U$ could be obtained by analyzing the transformation equations of $N$ and $T$, albeit $\left[N, T \right] \neq i$. It turns out that in the case of crystals there is a particularly useful set of operators $N,T$ of this type:

\bea
N = \frac{x}{a}, \qquad T = \exp \left(- iap \right),
\label{5}
\eea
with the algebraic properties

\bea
\left[ N, T \right] = T, \qquad T T^{\dagger} = T^{\dagger}T= \v 1.
\label{5.1}
\eea
These operators have a simple interpretation in terms of lattice coordinates, as they act on atomic states in the form
\bea
N | n \> = n | n \>, \qquad T | n \> = | n + 1 \>. 
\label{6}
\eea
By means of (\ref{5}) we may infer that any unitary transformation of $x,p$ leads again to $N' = U N U^{\dagger}$,  $T' = U T U^{\dagger}$, ensuring the covariance of the commutator $\left[ N', T'\right]=T'$. Furthermore, the relations in (\ref{6}) show that the matrix elements of these operators are considerably simpler than those of $\xi$ and $\pi$:

\bea
N_{n,m} = n \delta_{n,m}, \qquad T_{n,m} = \delta_{n,m+1}.
\label{6.1}
\eea
Finally, our study of the representations $\<x|U|x'\>$ can be reduced to the analysis of $U_{n,m} \equiv \<n|U|m\>$.

\subsection{Non-invertibility and ambiguity groups}

The transformation equations (\ref{1}), (\ref{5}) and the property $T^{\dagger} = 1/T$ imply that the map $(N,T) \mapsto (\xi, \pi)$ can be written explicitly as

\bea
\xi (N,T) = a \left\{\left( T + T^{\dagger} \right)^{-1}, N \right\}= a \sum_{m=0}^{\infty}(-)^m \left( 2\left[ N-m\right] + 1 \right) T^{2m+1} , \nonumber \\ \pi (T) = \frac{1}{2ia} \left( T + T^{\dagger} \right)= \frac{1}{2ia} \left( T + \frac{1}{T} \right). 
\label{7}
\eea
This map cannot be inverted for $N$ and $T$, since it is 2 to 1. Moreover, the transformation (\ref{5}) relating $(N,T)$ with the original variables $(x,p)$, maps an infinite number of Riemann sheets to one plane. According to \cite{moshLASP} and references cited therein, this gives rise to what is known as {\it ambiguity spin.\ }In other words, the space representation of $U$ is multiply defined by all matrix elements of the type $\<x| U | x'\> \exp \left[ 2 \pi i q (x-x')/a \right]$, where $q \in \v Z$. This can be seen more clearly by recognizing that in the space $(x,p)$ all the preimages of the form

\bea
p_q \equiv p+\frac{2\pi q}{a} = \exp \left[ -i\frac{2\pi q x}{a} \right] \, p \, \exp \left[ i\frac{2\pi q x}{a} \right]
\label{8}
\eea
lead to the same image $(N,T)$. An infinite dimensional matrix with spinorial indices $q,q'$ should take care of the ambiguity and it can be proposed in the form

\bea
U^{q,q'}_{n,m} = \int \int dx dx' \psi_n^*(x) \psi_m(x') \<x| U | x'\> \times \exp \left[ i\frac{2\pi (q x- q' x')}{a} \right] \delta_{q,q'}.
\label{9}
\eea
Such $U^{q,q'}_{n,m}$ represents a transformation between $(N,T)$ and $(N',T')$ in discrete coordinates $n$ and $m$, just as $\<x| U |x'\>$ represents the corresponding transformation between $(x,p)$ and $(x',p')$ in the Hilbert space $\{\psi_n(x) \}$. The ambiguity of the crystalline group, \ie $( \v Z, + )$, is therefore represented irreducibly by $ \exp \left[ i\frac{2\pi  q(x- x')}{a} \right] \delta_{q,q'}$. 

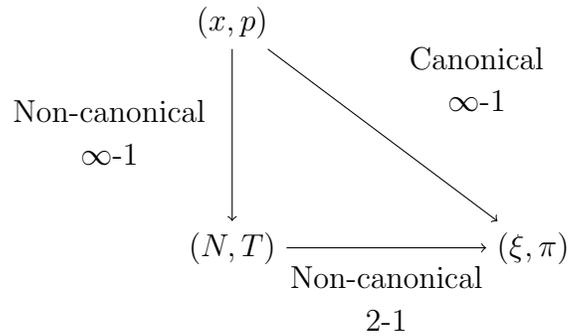
\begin{figure}

\begin{center}

\begin{tikzpicture}[node distance=3cm, auto] 

\node (a) {$(x,p)$};

\node (b) [below of=a] {$(N,T)$};

\node (c) [right of=b,node distance=4cm] {$(\xi, \pi)$};

\draw[->] (a) to node {\begin{tabular}{c} Canonical \\ $\infty$-$1$ \end{tabular}} (c);

\draw[->] (a) to node [swap]  {\begin{tabular}{c} Non-canonical \\ $\infty$-$1$ \end{tabular}} (b);

\draw[->] (b) to node [swap] {\begin{tabular}{c} Non-canonical \\ $2$-$1$ \end{tabular}}  (c);

\end{tikzpicture} 

\end{center}
\caption{Commutative diagram showing the maps (\ref{1}), (\ref{5}) and (\ref{7}).}
\label{diag1}
\end{figure}

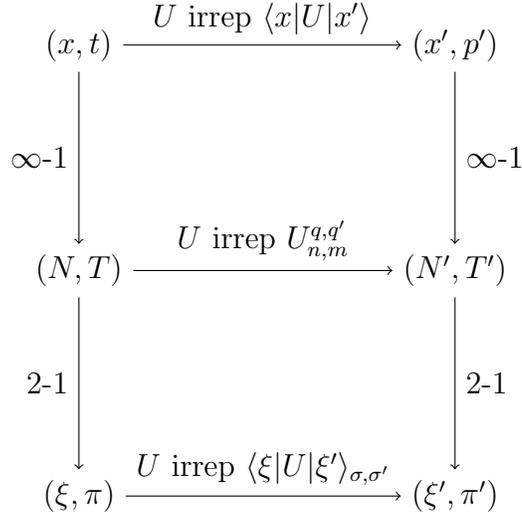
\begin{figure}

\begin{center}

\begin{tikzpicture}[node distance=3cm, auto]

\node (A) {$(x,t)$};

\node (A') [right of=A, node distance=5cm] {$(x',p')$};

\node (B) [below of=A] {$(N,T)$};

\node (B') [below of=A'] {$(N',T')$};

\node (C) [below of=B] {$(\xi,\pi)$};

\node (C') [below of=B'] {$(\xi',\pi')$};

\draw[->] (A) to node {$U$ irrep $\<x|U|x'\>$} (A');

\draw[->] (B) to node {$U$ irrep $U^{q,q'}_{n,m}$} (B');

\draw[->] (C) to node {$U$ irrep $\<\xi|U|\xi'\>_{\sigma,\sigma'}$} (C');

\draw[->] (A) to node [swap] {$\infty$-$1$} (B);

\draw[->] (B) to node [swap]  {$2$-$1$} (C);

\draw[->] (A') to node {$\infty$-$1$} (B');

\draw[->] (B') to node  {$2$-$1$} (C');

\end{tikzpicture} 

\end{center}
\caption{Diagram showing the canonical transformation $U$ and its Hilbert space representations, for each pair of variables. In this work we study the map $N'=UNU^{\dagger}$ $T'=UTU^{\dagger}$ represented by $U^{q,q'}_{n,m}$. The third level contains extra indices $\sigma=\pm$, $\sigma'=\pm$ due to a $2$-$1$ map.}
\label{diag2}
\end{figure}

To summarize, we have three maps relating the pairs $(x, p)$, $(\xi, \pi)$ and $(N, T)$. The diagram in fig.\ref{diag1} commutes, but the morphisms are not necessarily one-to-one. The ambiguity spin comes to our rescue, and in fact it is possible to work with the representation $U^{q,q'}_{n,m}$ alone. The three levels at which the canonical transformation is represented in Hilbert space are indicated in the diagram of fig.\ref{diag2}.

In the rest of this paper we shall work with $q=0$. We recognize that such a choice is equivalent to a particular phase factor of the (tight-binding) functions $\psi_n(x)$; the r.h.s. of (\ref{9}) shows it clearly, for the replacement $\psi_n(x) \mapsto e^{i 2\pi q x/a} \psi_n(x) $ does the required job.

Ultimately, the MM equations for a canonical transformation can be written for any pair of independent variables $(N,T)$ \cite{moshLASP} and we shall proceed in this direction in what follows.

\section{The discrete version of the MM equations}

\subsection{Transformation equations with a continuous parameter}

In full analogy with what is done for the harmonic oscillator and the associated symplectic group \cite{moshandquesne}, let us consider a set of transformation equations $N'(N,T)$ and $T'(N,T)$ which represent the evolution of the operators $(N,T)$ under some hamiltonian.

Since the pair $(N,T)$ is now our main focus, it should be stressed that if $T^{\dagger} = 1/T$, \ie if $T^{\dagger}$ is not independent of $T$, the commutator $\left[ N, T^{\dagger} \right] = -T^{\dagger}$ is redundant. With this in mind, we propose the non-linear transformation

\bea
N' = \alpha \, T + \frac{\alpha^*} {T} + \beta N,
\label{10}
\eea 
\bea
T' = \gamma \, T + \frac{\bar \gamma} {T} + \delta N,
\label{11}
\eea 
where $\alpha, \gamma, \bar \gamma$ and $\delta$ are complex functions of a continuous parameter $t$ (time) and $\beta$ is real. The conditions $T' T'^{\dagger} = T'^{\dagger}T' = \v 1$ and $[N',T']=T'$ lead to restrictions on such functions, \ie

\bea
\gamma \beta - \alpha \delta= 1, \qquad \alpha^* \delta = \bar \gamma \beta, \qquad
\delta \gamma^* = \delta^* \gamma, \qquad \delta \bar \gamma^* = \delta^* \bar \gamma.
\label{12}
\eea
From (\ref{10}), (\ref{11}) and the unitary operator $U$ acting from the right, we have

\bea
U N = \alpha \, T U + \alpha^* T^{\dagger} U + \beta N U,
\label{13}
\eea 
\bea
U T = \gamma \, T U + \bar \gamma T^{\dagger} U + \delta N U.
\label{14}
\eea 
Finally, the matrix elements of both relations (\ref{13}), (\ref{14}) can be found using the atomic states $\< n |$, $|m \>$ and (\ref{6.1}) :

\bea
m U_{n,m}  = \alpha \, U_{n-1,m} + \alpha^*  U_{n+1,m} + \beta n U_{n,m},
\label{15}
\eea 
\bea
U_{n,m+1} = \gamma \, U_{n-1,m} + \bar \gamma U_{n+1,m} + \delta n U_{n,m}.
\label{16}
\eea 
These are two coupled recurrence relations in the indices $n,m$ and their solution is the unitary representation of the canonical transformation in discrete coordinates. We shall refer to it as the discrete MM equations. Specific choices of the coefficients $\alpha, \beta, \gamma, \delta, \bar \gamma$ yield solutions in terms of Bessel functions \cite{5, 5.1}. A concrete example is given in the following section.

\subsection{The Stark Hamiltonian}

Suppose now that a concrete hamiltonian is given. For a particle in discrete space with a time-dependent hopping parameter $\alpha$ and under the influence of a time-dependent homogeneous field $\beta$, we have

\bea
H = \alpha(t) T + \alpha^*(t) T^{\dagger} + \beta(t) N.
\label{17}
\eea
A canonical transformation is obtained by means of the unitary evolution, therefore we analyze the equations of motion for $N$ and $T$:

\bea
\dot T = i \beta(t) T, \qquad \dot T^{\dagger} = -i \beta(t) T^{\dagger}, \qquad \dot N = i \alpha^{*}(t) T^{\dagger} - i \alpha(t) T. 
\label{18}
\eea
The solutions of (\ref{18}) yield the evolution map, which results in

\bea
\fl T(t) = e^{i f(t)} T(0), \qquad T^{\dagger}(t) = e^{-i f(t)} T^{\dagger}(0), \qquad N(t) =  F^*(t) T^{\dagger} + F(t)  T +  N(0), 
\label{19}
\eea
where we have used the definitions $f(t) = \int^{t} \beta(t) dt$ and $F(t) = -\int^{t} i \alpha(t) e^{i f(t)} dt $. The discrete MM equations corresponding to such a map are a particular case of (\ref{15}) and (\ref{16}) in the previous section. By computing the corresponding matrix elements, we obtain the recurrences

\bea
(m-n) U_{n,m}  = F(t)  U_{n-1,m} + F^*(t)  U_{n+1,m},
\label{20}
\eea 
\bea
U_{n,m+1} =  e^{i f(t)} \,  U_{n-1,m},
\label{21}
\eea 
and their solutions are explicitly given by

\bea
U_{n,m}(t) = e^{imf(t)} \left[ \frac{F(t)}{F^*(t)} \right]^{(n-m)/2} J_{m-n} \left( 2 |F(t)| \right).
\label{22}
\eea
Here $J$ denotes a Bessel function of the first kind; the Bessel function of the second kind $Y$ is not allowed, as it does not fulfill the requirement $U_{n,m}(0) = \delta_{n,m}$ due to its irregularity at the origin.

\section{The MM equations in two-dimensional lattices and their solution}

The next non-trivial step is to increase the number of dimensions. For crystalline sheets with tight-binding hamiltonians and a time-dependent field, there are two examples that offer themselves: the square lattice and the triangular lattice. We shall see that the topological differences between these tessellations lead to different canonical maps; therefore, the corresponding propagators are different.

\subsection{Square lattice immersed in an external field}

Let $T_i = \exp\left( -i a_i p_i\right)$ be the translation operator along the $i$-th primitive vector ($i=1$ for \v i and $i=2$ for \v j) with lattice spacings $a_i$ for each direction. For a field with components $\beta_i(t)$, the hamiltonian of interest has the form

\bea
H = \alpha_1(t) T_1 + \alpha_2(t) T_2 + \mbox{h.c.} + \beta_1(t) N_1 + \beta_2(t) N_2.
\label{23}
\eea
At this point it is convenient to introduce an appropriate Hilbert space: we denote the atomic states by $|n_1, n_2 \>$ where $n_1, n_2$ are integers. We have 

\bea
N_1 |n_1, n_2 \> = n_1 |n_1, n_2 \>, \qquad N_2 |n_1, n_2 \> = n_2 |n_1, n_2 \>, \\
T_1 |n_1,n_2 \> = |n_1+1,n_2 \>, \qquad T_2 |n_1, n_2 \> = |n_1,n_2+1 \>.
\eea
It is important to note that $H$ can be separated in two parts which commute: a part which contains operators with index 1 and another with index 2. The coordination number $c=4$ makes possible a separation of variables and we expect that the resulting MM equations can be decoupled in two sets of independent equations. In fact, we have

\bea
(m_1-n_1) U_{n_1,m_1;n_2,m_2}  = F_1(t)  U_{n_1-1,m_1;n_2,m_2} + F_1^*(t)  U_{n_1+1,m_1;n_2,m_2}, \\
(m_2-n_2) U_{n_1,m_1;n_2,m_2}  = F_2(t)  U_{n_1,m_1;n_2-1,m_2} + F_2^*(t)  U_{n_1,m_1;n_2+1,m_2},
\label{24}
\eea 
\bea
U_{n_1,m_1+1;n_2,m_2} =  e^{i f_1(t)} \,  U_{n_1-1,m_1;n_2,m_2}, \\
U_{n_1,m_1;n_2,m_2+1} =  e^{i f_2(t)} \,  U_{n_1,m_1;n_2-1,m_2},
\label{25}
\eea 
which are easily solved by

\bea
\fl U_{n_1,m_1;n_2,m_2}(t) = \prod_{j=1,2} e^{im_jf_j(t)} \left[ \frac{F_j(t)}{F_j^*(t)} \right]^{(n_j-m_j)/2} J_{m_j-n_j} \left( 2 |F_j(t)| \right).
\label{26}
\eea
The parameters are given by $f_j(t)=\int^{t}\beta_j(t)dt$ and $F_j(t)=-\int^{t}i\alpha_j(t) e^{i f_j(t)}dt$.

\subsection{Triangular lattice immersed in an external field}

This example is crucial in our understanding of lattices. Although the corresponding MM equations can be solved, we shall see that their solutions cannot be expressed as a simple product. Take primitive lattice vectors $\v a_1 = \frac{1}{2}(\sqrt{3} \v i + 3 \v j) $, $\v a_2 =- \sqrt{3} \v i$ and $\v a_3 = -\v a_1 - \v a_2$, such that any lattice point can be expressed as $\v A = n_1 \v a_1 + n_2 \v a_2$. The translation operators are given by $T_i = \exp\left( -i \v a_i \cdot \v p_i\right)$ for each direction, with the additional property $T_3 = T^{\dagger}_1 T^{\dagger}_2$. With this notation, a hamiltonian analogous to (\ref{23}) but with coordination number $c=6$ can be proposed:

\bea
H= \sum_{i=1}^{3} \alpha_i (t) T_i + \mbox{h.c.} + \beta_1(t) N_1 + \beta_2(t) N_2.
\label{27}
\eea
The action of the operators $N_i$, $T_i$ on the atomic states $|n_1, n_2 \>$ is, as before,

\bea
N_1 |n_1, n_2 \> = n_1 |n_1, n_2 \>, \qquad N_2 |n_1, n_2 \> = n_2 |n_1, n_2 \>, \\
T_1 |n_1,n_2 \> = |n_1+1,n_2 \>, \qquad T_2 |n_1, n_2 \> = |n_1,n_2+1 \>.
\label{28}
\eea
However, in the triangular case we have the additional relations

\bea
T_3  |n_1,n_2 \> = |n_1-1,n_2-1 \>, \qquad \left[ T_3, N_j \right] = T_3, \quad j=1,2.
\label{29}
\eea
Let us define $\beta_3 = -\beta_1-\beta_2$ and $f_i(t) =\int^{t}\beta_i(t)dt$ for $i=1,2,3$. Similarly, take $F_j = -\int i \alpha_j(t) e^{i f_j(t)}dt$ for $j=1,2$, but define the third function as $F_3 = +\int i \alpha_3(t) e^{i f_3(t)}dt$, \ie with a reversed sign. With these definitions, the dynamical map generated by $H$ can be written in compact form as

\bea
T_i(t) = e^{i f_i(t)}T_i(0), \quad i=1,2,3. 
\label{30}
\eea
\bea
N_j(t) = F_j (t) T_j (0) + F_3(t)T_3(0) + \mbox{h.c.} + N_j(0), \quad j=1,2.
\label{31}
\eea
The relation (\ref{31}) differs considerably from the case of a square lattice. The discrete MM equations are now

\bea
\fl (m_1-n_1) U_{n_1,m_1;n_2,m_2}  &=& F_1(t)  U_{n_1-1,m_1;n_2,m_2} + F_1^*(t)  U_{n_1+1,m_1;n_2,m_2}, \nonumber \\
\fl &+& F_3(t)  U_{n_1+1,m_1;n_2+1,m_2} + F_3^*(t)  U_{n_1-1,m_1;n_2-1,m_2},
\label{31.1}
\eea
\bea
\fl (m_2-n_2) U_{n_1,m_1;n_2,m_2}  &=& F_2(t)  U_{n_1,m_1;n_2-1,m_2} + F_2^*(t)  U_{n_1,m_1;n_2+1,m_2}, \nonumber \\
\fl &+& F_3(t)  U_{n_1+1,m_1;n_2+1,m_2} + F_3^*(t)  U_{n_1-1,m_1;n_2-1,m_2},
\label{32}
\eea 
and
\bea
U_{n_1,m_1+1;n_2,m_2} =  e^{i f_1(t)} \,  U_{n_1-1,m_1;n_2,m_2},
\label{32.1}
\eea
\bea
U_{n_1,m_1;n_2,m_2+1} =  e^{i f_2(t)} \,  U_{n_1,m_1;n_2-1,m_2},
\label{33}
\eea 
where we have omitted the recurrence arising from $T_3(t) = e^{if_3(t)} T_3(0)$, which can be proved redundant. We may attempt a separable solution of the form

\bea
\fl U_{n_1,m_1;n_2,m_2} = \exp \left[i\frac{(n_1+m_1)f_1(t)}{2} +  i\frac{(n_2+m_2)f_2(t)}{2} \right] \times K_{m_1-n_1,m_2-n_2}(t),
\label{34}
\eea
which solves (\ref{32.1}) and (\ref{33}). Substitution of (\ref{34}) in (\ref{31.1}) and (\ref{32}) gives the simplified recurrences

\bea
\fl n K_{n,m} = F_1 e^{-i\frac{f_1}{2}} K_{n+1,m} + F^*_1 e^{i\frac{f_1}{2}} K_{n-1,m} +  F_3 e^{-i\frac{f_3}{2}} K_{n-1,m-1} + F^*_3 e^{i\frac{f_3}{2}} K_{n+1,m+1},
\label{35}
\eea
\bea
\fl m K_{n,m} = F_2 e^{-i\frac{f_2}{2}} K_{n,m+1} + F^*_2 e^{i\frac{f_2}{2}} K_{n,m-1} +  F_3 e^{-i\frac{f_3}{2}} K_{n-1,m-1} + F^*_3 e^{i\frac{f_3}{2}} K_{n+1,m+1}.
\label{35.1}
\eea
Therefore, we have reduced a problem of four variables $n_1,n_2,m_1,m_2$ to two variables $n = m_1-n_1$ and $m=m_2-n_2$. The two recurrence relations (\ref{35}), (\ref{35.1}) are indeed simple; they can be solved by generalizations of Bessel functions in full analogy with the recurrence (\ref{20}), but now extended to two indices and three variables. One of such generalizations is of particular interest, as it was studied in \cite{1.3} in a completely different context. Suppose now that the parameters in (\ref{35}) are such that 

\bea
\mbox{arg} (F_3) = \mbox{arg} (F_1) + \mbox{arg} (F_2).
\label{36}
\eea  
This restriction of the parameters is not completely unnatural, as it includes the free problem in the triangular lattice and the problem of equal couplings $\alpha_1=\alpha_2=\alpha_3$ and constant external field $\dot \beta_i = 0$, among other interesting possibilities. Therefore we may use (\ref{36}) to further reduce (\ref{35}) and (\ref{35.1}), arriving at the following relations

\bea
\fl n G_{n,m} = |F_1| \left( G_{n+1,m} +  G_{n-1,m} \right) +  |F_3| \left( G_{n-1,m-1} + G_{n+1,m+1} \right), \\
\fl m G_{n,m} = |F_2| \left(  G_{n,m+1} + G_{n,m-1} \right) +  |F_3| \left( G_{n-1,m-1} + G_{n+1,m+1} \right).
\label{37}
\eea
These relations are uniquely solved (recall our intial condition $U(0)=\v 1$) by Bessel functions of two indices, \ie $G_{n,m}(t)= J^{(+,-)}_{n,m}\left(2|F_1(t)|,2|F_2(t)|,2|F_3(t)| \right)$, which is a special case of (14) in \cite{1.3}. Thus, the full solution for our unitary operator reads

\bea
\fl U_{n_1,m_1+1;n_2,m_2} &=& \exp \left[ i\frac{(n_1+m_1)f_1(t)}{2} +  i\frac{(n_2+m_2)f_2(t)}{2} \right]  \nonumber \\
\fl &\times&  \exp \left[ i(m_1-n_1) \Phi_1 + i (m_2-n_2) \Phi_2 \right] \nonumber \\ &\times& J^{(+,-)}_{m_1-n_1,m_2-n_2}\left(2|F_1(t)|,2|F_2(t)|,2|F_3(t)| \right),
\label{38}
\eea
with the abbreviation $\Phi_j = f_j(t)/2 - \mbox{arg}[F_j(t)] $. 

It is important to note that the two-index, three-parameter Bessel function $J^{(+,-)}_{n,m}$ comes from an exponential generating function of the type 

\bea
e^{\left[ x(a - 1/a)+y(b - 1/b)+z(ab - 1/ab) \right]/2} = \sum_{n,m \in \v Z} a^n b^m J^{(+,-)}_{n,m}(x,y,z).
\label{39}
\eea
This simple origin of our special functions brings forth a number of properties and relations that can be obtained for our propagator (\ref{38}). In particular, the partial derivatives of $J$ with respect to $x,y,z$ can be employed to show that (\ref{38}) is indeed the Green's function of the time-dependent Schr\"odinger equation with hamiltonian (\ref{27}). Other properties such as expansions in products of ordinary Bessel functions and ascending series in $x,y,z$ may be helpful in the numerical evaluation of the kernel.

\section{The discrete MM equations for arbitrary lattices}

Now that we have solved two relevant examples, we proceed to generalize our treatment to many dimensions and coordination numbers. Although we do not solve explicitly the resulting relations for the general case, we may sketch the path towards a reduction of the problem. Consider a set of translation operators $T_i = \exp \left(-i \v a_i \cdot \v p \right)$ along lattice vectors $\{\v a_i \}_{i=1,...,c}$ with $c$ the coordination number. Let the primitive vectors of the lattice be the subset $\{\v a_j\}_{j=1,...,d}$ with $d$ the dimension of the space; usually we have $d \leq c$. Finally, let us denote by $\{N_j\}_{j=1,...,d}$ the operators of discrete positions along the primitive vectors $a_j$.

With this notation, consider now the hamiltonian

\bea
H= \sum_{i=1}^{c} \alpha_i(t) T_i + \mbox{h.c.} + \sum_{j=1}^{d} \beta_j(t) N_j.
\label{40}
\eea
The evolution map on the operators $N_j, T_i$ can be obtained by virtue of the commutation relations

\bea
\left[ N_j , T_i \right] = \delta_{ji} T_i, \quad \mbox{if} \quad i \leq d
\label{41}
\eea
and 
\bea
\left[ N_j , T_l \right] = q_l T_l, \quad \mbox{for some} \quad l > d.
\label{42}
\eea
The last relation is clearly determined by the topology of the lattice, for which some of the neighbouring sites can be connected only by linear combinations of the primitive vectors and not by the primitives themselves -- the triangular lattice is a clear example with $q_3=-1$. We may encode both (\ref{41}) and (\ref{42}) in a function $\sigma$ for simplicity:

\bea
\left[ N_j , T_i \right] = \sigma_{ji} T_i, \quad  1 \leq j \leq d, \, 1 \leq i \leq c.
\label{43}
\eea
From this commutator, it is straightforward to find the Heisenberg equations of motion and extract the dynamical map. We obtain 

\bea
T_i(t) = e^{i f_i (t)} T_i(0), \\
N_j(t) = N_j(0) + \sum_{i=1}^{c} F_{ji}(t) T_i (0) + F_{ji}^*(t) T^{\dagger}_i (0).
\label{44}
\eea
The definition of parameters is now

\bea
f_i(t)= \sum_{j=1}^d \sigma_{ji} \int_0^{t} \beta_j(t) dt , \qquad F_{ji} (t) = -i \sigma_{ji} \int_0^{t} \alpha_i(t) e^{i f_i(t)} dt.
\label{45}
\eea
At this stage it is convenient to introduce the atomic states of the lattice $| \{ m_i \}_{i=1,...,d}\>$ and $\< \{ n_i \}_{i=1,...,d}|$. With these states, we are able to compute the matrix elements of (\ref{44}) and finally write down the corresponding MM equations:

\bea
\fl U_{n_1...n_d; m_1...m_i+1...m_d} = e^{i f_i (t)} U_{n_1...n_i-1...n_d; m_1...m_d},
\label{45.1}
\eea
\bea
\fl (m_j-n_j) U_{n_1...n_d; m_1...m_d} = \sum_{i=1}^{c}\left[ F_{ji}(t) U_{n_1...n_i-1...n_d; m_1...m_d} + F_{ji}^*(t) U_{n_1...n_i+1...n_d; m_1...m_d} \right],
\label{46}
\eea
which is a set of $d$ recurrences. As before, we have omitted a number $c-d$ of redundant equations arising from $T_l(t), l=d,...,c$. Repeating our previous steps for low-dimensional problems, the first equation can be directly solved by products of the type

\bea
U_{n_1...n_d;m_1...m_d} = \exp\left( \frac{i}{2} \sum_{j=1}^{d} (n_j+m_j) f_j(t)  \right) K_{m_1-n_1,...,m_d-n_d}(t).
\label{47}
\eea 
The final relation for $K$ becomes a set of $d$ recurrences for which no precedent can be found in the literature. It seems plausible that new generalizations of Bessel functions may arise from such equations. In all, we can say that the representations of a certain type of canonical transformations in arbitrary lattices have been characterized as the solutions of a specific finite-difference equation of many indices, namely the relations (\ref{45.1}) and (\ref{46}).  

\section{Conclusion}

Our efforts have led us to the construction of discrete representations of canonical transformations and their explicit forms by means of dynamical maps. In the first part we developed the necessary techniques in one-dimensional lattices, finding thereby that the propagator of a particle in discrete space with time-dependent site couplings and a time-modulated homogeneous field can be written in terms of Bessel functions. In full analogy with such a simple example, we went on and obtained two new propagators for two-dimensional lattices immersed in a time-dependent field. Finally, we indicated the path to follow in the case of multidimensional lattices of an arbitrary topology. 

With this, it is fair to say that particles in crystals (in the basis of Wannier functions) are susceptible of phase space analysis, and that a number of relevant models can be solved explicitly in time domain.

\ack

I am grateful to A. Turbiner for bringing his work to my attention. Financial support from PROMEP Project $103.5/12/4367$ is acknowledged.

\end{document}